# On the configurational force associated with blocked slip bands at grain boundaries in α-Ti

Abdalrhaman Koko [1,2]*

[1] National Physical Laboratory, Hampton Road, Teddington TW11 0LW, UK

[2] Department of Materials Science, University of Oxford, Parks Road, Oxford OX1 3PH, UK

* Corresponding author: abdo.koko@npl.co.uk

## Abstract

Grain boundaries can block slip-band propagation and generate intense local stress and strain fields that influence subsequent deformation and damage initiation in polycrystalline metals. Conventional geometric criteria, such as Schmid factor and slip-transfer parameters, describe crystallographic compatibility but do not quantify the energetic severity of a blocked slip event. Here, we apply a configurational force framework to high-angular-resolution electron backscatter diffraction (HR-EBSD) measurements obtained from a blocked slip band in commercially pure titanium. By evaluating a *J*-type equivalent domain integral from the measured elastic field, we quantify both the magnitude and directional dependence of the local energetic driving force associated with the stress localisation; thus, providing an energetic descriptor of the tendency for deformation to extend into the neighbouring grain. The results show a marked decoupling between conventional geometric metrics and the configurational force response, indicating that the local stress-localisation geometry strongly influences which crystallographically admissible extension directions in the neighbouring grain are energetically favoured. The framework provides a physically grounded basis for quantifying blocked-slip severity and for motivating future in situ studies aimed at defining a critical transfer threshold for transfer or cracking.

Keywords: HR-EBSD; Slip transfer; Grain boundary; Configurational force; Titanium

# 1. Introduction

Grain boundaries can interrupt slip-band propagation, critically influencing plasticity and strengthening in polycrystalline metals [1–3]. Depending on crystallographic alignment and boundary character, a slip band may transmit across the boundary, generate an intense stress concentration, or be fully arrested [4–6]. When a slip band impinges on a grain boundary, dislocations accumulate and form a pile-up that generates highly localised stress and strain fields [7–10]. The resulting stress concentration ahead of the blocked band may activate deformation in the neighbouring grain, including slip [11,12] or twinning [13,14], or may remain confined to the boundary region and promote cavity or crack initiation [15,16].

Classical descriptors such as Schmid factor, residual Burgers vector, or the Lustre–Morris ($m'$) parameter capture crystallographic compatibility but do not quantify the magnitude of the driving force, as high geometric alignment alone is insufficient for transmission unless adequate stress and elastic energy accumulate to overcome the boundary resistance. Advances in high-angular-resolution electron backscatter diffraction (HR-EBSD) have enabled quantitative mapping of the elastic displacement gradient (strain and rotation) and stress fields surrounding blocked slip bands with sub-micron resolution [17,18]. These measurements have provided direct experimental validation of classic dislocation pile-up models by revealing singular stress fields ahead of impinging slip bands and confirming the expected inverse square-root dependence of resolved shear stress with distance from the boundary [19–21]. By projecting the elastic stress onto the incoming slip system, previous studies have extracted effective Hall–Petch-type coefficients and demonstrated the importance of boundary geometry and orientation in resisting slip [22–25].

Despite these advances, the current state of the art remains limited by its reliance on single-stress proxies, with quantitative analyses of blocked slip bands often reducing the full deformation field to a fitted resolved shear stress on the incoming slip plane. Such approaches do not capture the total energetic severity of a blocked slip event, nor do they quantify how much elastic energy is transmitted into the neighbouring grain. Here we show that configurational (Eshelby) force [26,27], evaluated directly from HR-EBSD-measured elastic fields, provides a compact energetic descriptor of blocked slip bands that cannot be inferred from geometric metrics alone, while using the full stress and strain tensor rather than a single fitted component.

# 2. Methodology

We analyse the CP-Ti dataset of Guo et al. [19], acquired at 20 kV (14 nA) with 0.2 μm step size after ~1% tensile strain. ~1000 × 1000-pixel EBSD patterns were recorded and cross-correlated against an in-grain reference far from the band/GB. A two-pass correlation procedure was employed [28,29], with 50 regions of interest (256 × 256 pixels). The first pass estimated rigid-body lattice rotations and remapped patterns; the second resolved elastic displacement gradients ($u_{i,j}$) without rigid-body components, and infinitesimal elastic strains, $\varepsilon_{ij} = \frac{1}{2}\left(u_{i,j} + u_{j,i}\right)$. Anisotropic elastic constants (i.e., $C_{11}$ = 162.4, $C_{33}$ = 180.7, $C_{44}$ = 177.0, $C_{66}$ = 35.2, $C_{12}$ = 92.0, and $C_{13}$ = 69.0 in GPa ) were used to convert measured elastic strains to stresses ($\sigma_{ij}$) into the crystal reference frame [30], while assuming plane-stress condition normal to the free surface ($\sigma_{33} = 0$ [31]).

The configurational (Eshelby) force associated with slip-band localisation induced by a blocked slip band in the neighbouring grain (Grain B) can be derived starting from Eshelby's definition of the energy-momentum tensor ($P_{kj}$) [26,27], which expresses the configurational forces acting on defects in an elastic body:

$$P_{kj} = W \delta_{jk} - \sigma_{ij} u_{i,k}, \qquad k = 1,2$$

$$\because W = \frac{1}{2}\sigma_{ij}\,\varepsilon_{ij}, \qquad u_{i,k} = \frac{\partial u_i}{\partial x_k} \qquad\qquad 1$$

where $W$ is the strain energy density and $\delta_{jk}$ is the Kronecker delta. The configurational force ($F_k$) acting on a defect in the $x_k$ direction is obtained by integrating the divergence of $P_{kj}$ over a contour ($\Gamma$) surrounding the defect [32]:

$$F_k = \int_{\Gamma} P_{kj} n_j \, d\Gamma \qquad\qquad 2$$

where $n$ is the components of the unit vector normal to contour. For a virtual extension along $x_1$, the relevant component can be expressed as a path-independent contour integral:

$$F_1 = \int_\Gamma \left( W\,\delta_{j1} - \sigma_{ij}u_{i,1} \right) n_j\,d\Gamma \qquad\qquad 3$$

Equation (3) describes the Eshelby configurational force acting on a defect (resisting movement), but here we want the energy available from extension (energy release) available from the stress localisation. This changes the formulation to equation (4), which is similar to the *J*-type integral formulation [33]. For computational implementation, the contour integral is transformed into an equivalent domain integral using a weighting function ($q$), leading to the equivalent domain integral (EDI) formulation (equation 4) [34,35], thereby reducing sensitivity to mesh refinement and numerical noise [36,37] and allowing us to calculate the localisation's configurational force with the integral projected along a prescribed virtual extension direction (VED) (Figure 1) [38]. Given that HR-EBSD displacement fields were acquired on a regular square grid, the evaluation of the configurational force along a given VED did not require interpolation or shape functions to remap measurement points. Instead, the integration domain ($dA$) was expanded incrementally in steps of 0.2 µm, corresponding to the HR-EBSD step size, around the region of intense stress concentration in Grain B until convergence of the integral was achieved. Hence, the formulation becomes:

$$J = -F_1 = \int_A \left( \sigma_{ij}u_{i,1} - W\,\delta_{j1} \right) \frac{\partial q}{\partial x_j}\,dA \;\; = \left( \sum_{j=1}^{2} \left( \sum_{i=1}^{3} \sigma_{ij}u_{i,1} - W\delta_{j1} \right) \frac{\partial q}{\partial x_j} \right) dA \qquad 4$$

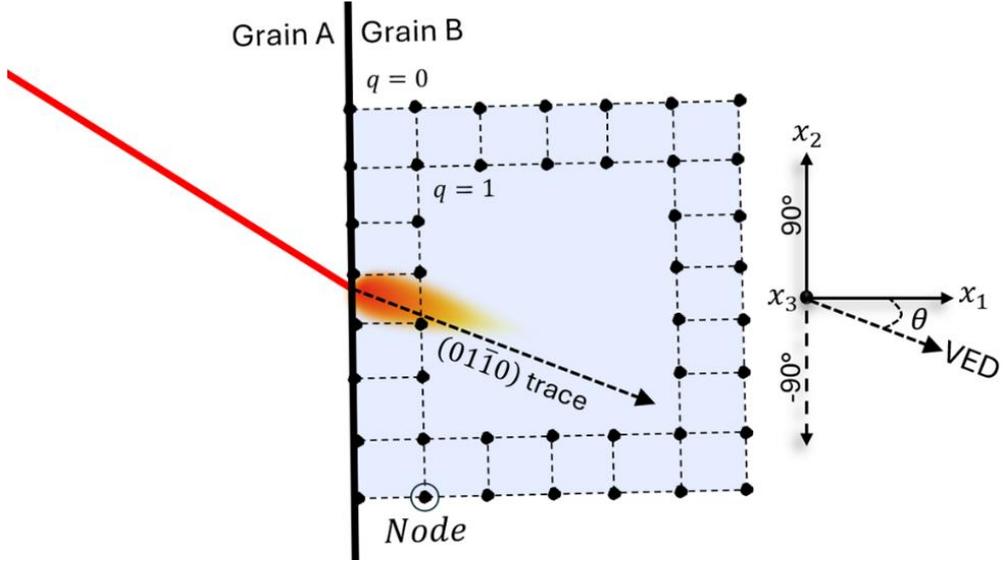

Figure 1: Schematic representation of a slip band intersecting a grain boundary between Grain A and Grain B. The red line indicates the slip band terminating at the boundary, with localised stress concentration shown as a gradient hotspot (red to yellow) at the intersection. The shaded blue square represents the integral domain for configurational force evaluation for the $(01\bar{1}0)$ slip trace. $q$ is a scalar function that equals 1 inside the inner domain and 0 outside the outer domain, ensuring smooth transition across the integration region.

To facilitate the configurational force analysis, the displacement gradient fields in the vicinity of the blocked slip band were first transformed into a consistent reference frame by swapping the 1- and 2-axes shown in Figure 2B, such that they aligned with the horizontal $x_2$ and $x_1$ axes of the analysis frame (Figure 1), respectively. The same sequence of transformations was applied to the elastic stiffness tensor to ensure consistency between the kinematic and constitutive descriptions [39].

For each evaluation, the VED was aligned with the $x_1$-axis (Figure 1). To achieve this, the displacement gradient tensors ($u_{i,j}$) were rotated into a local coordinate system aligned with the trace of each candidate slip variant in Grain B using a rotation matrix ($R_z(\theta)$, equation 5), where $\theta$ is the angle between the $i$th slip-trace direction and the $x_1$-axis. This procedure enables the calculation of the configurational force associated with each potential slip system in Grain B, rather than along the slip trace in Grain A.

$$u_{i,j}^r = R_z(\theta)\, u_{i,j}\, R_z^T(\theta), \qquad R_z(\theta) = \begin{bmatrix} \cos\theta & -\sin\theta & 0 \\ \sin\theta & \cos\theta & 0 \\ 0 & 0 & 1 \end{bmatrix} \qquad 5$$

Slip-trace analysis was conducted using EBSD-derived mean grain orientations and slip-system geometry implemented in the MTEX toolbox [40]. For each grain, the selected slip systems were transformed into the sample reference frame using the measured orientation, and slip-plane traces were calculated as the intersection of the slip planes with the sample surface. Symmetry-equivalent variants producing overlapping projected traces were consolidated to avoid redundant trace assignments. The 18 candidate slip variants were restricted to basal ⟨a⟩, prismatic ⟨a⟩, and first-order pyramidal ⟨c+a⟩ families, which are known to dominate plastic deformation in α-Ti under the applied loading conditions [41,42]. Higher-order slip systems and twinning modes were excluded due to their substantially higher critical resolved shear stresses [41–43], and the absence of experimental evidence for their activation in the analysed region. Thus, while multiple crystallographically distinct slip traces are commonly observed across grains, in Grain B, the projected slip systems collapse to four unique trace directions, with ± 1.2°, due to crystallographic symmetry and trace overlap (Figure 2B).

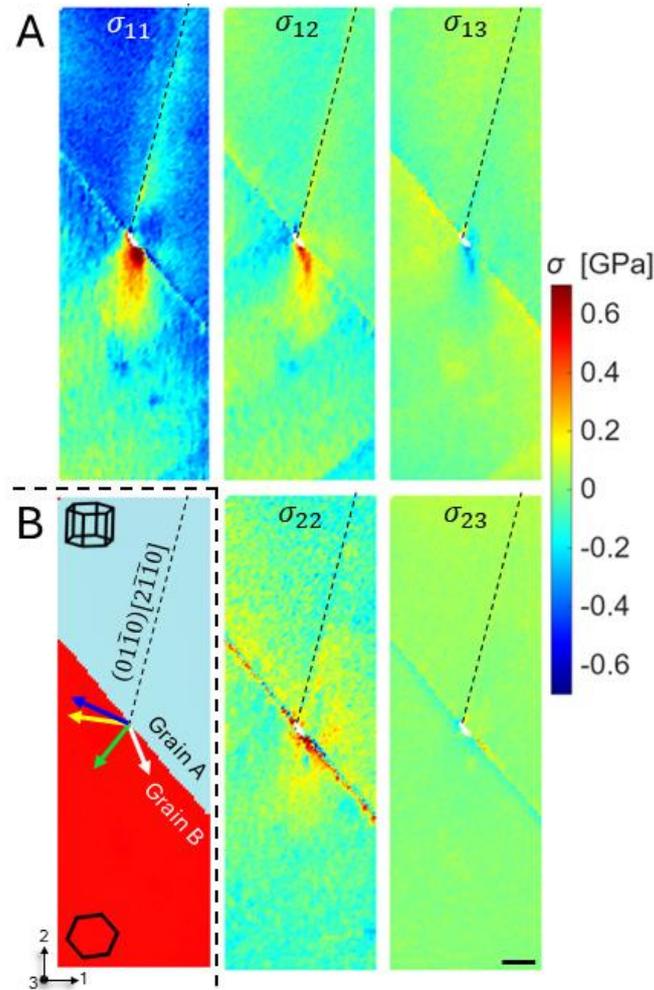

Figure 2: (A) Maps of the in-plane elastic stress components reconstructed from HR-EBSD–measured deviatoric elastic strains. The dashed line in each stress map indicates the blocked slip band trace. The scale bar in the $\sigma_{23}$ map corresponds to 2 µm. (B) EBSD orientation map of the same grain, showing the crystallographic orientation and the corresponding slip trace (dashed line). Coloured lines indicate candidate slip directions projected onto the sample surface.

# 3. Results and discussion

The termination of this slip band at the grain boundary inhibits further dislocation motion, leading to the accumulation of geometrically necessary dislocations and elevated elastic strain gradients. This incompatibility of plastic deformation across the boundary manifests as a localised stress concentration within the neighbouring grain (Grain B), as captured by the HR-EBSD–derived stress fields (Figure 2A). The calculated elastic stress components reveal a pronounced localisation at the grain boundary resulting from the impingement of a slip band originating in Grain A. The grain boundary misorientation was measured as 73.5 ± 0.3°, and the active slip band in Grain A is crystallographically consistent with the $(01\bar{1}0)[2\bar{1}\bar{1}0]$ prismatic ⟨a⟩ slip system within ± 0.8°. This system corresponds to the highest resolved shear stress in Grain A, with a Schmid factor of approximately 0.5.

When evaluated using the *J*-type domain integral, the calculated configurational force initially varies with integration-domain size but rapidly stabilises once the domain fully encloses the region of intense stress localisation adjacent to the grain boundary (Figure 3A), with residual sensitivity at small domain sizes arising from incomplete field capture and measurement noise in the HR-EBSD displacement gradients associated with severe deformation. Beyond this point, further domain expansion produces no systematic change in the evaluated force, confirming that the result is insensitive to domain size and local numerical noise.

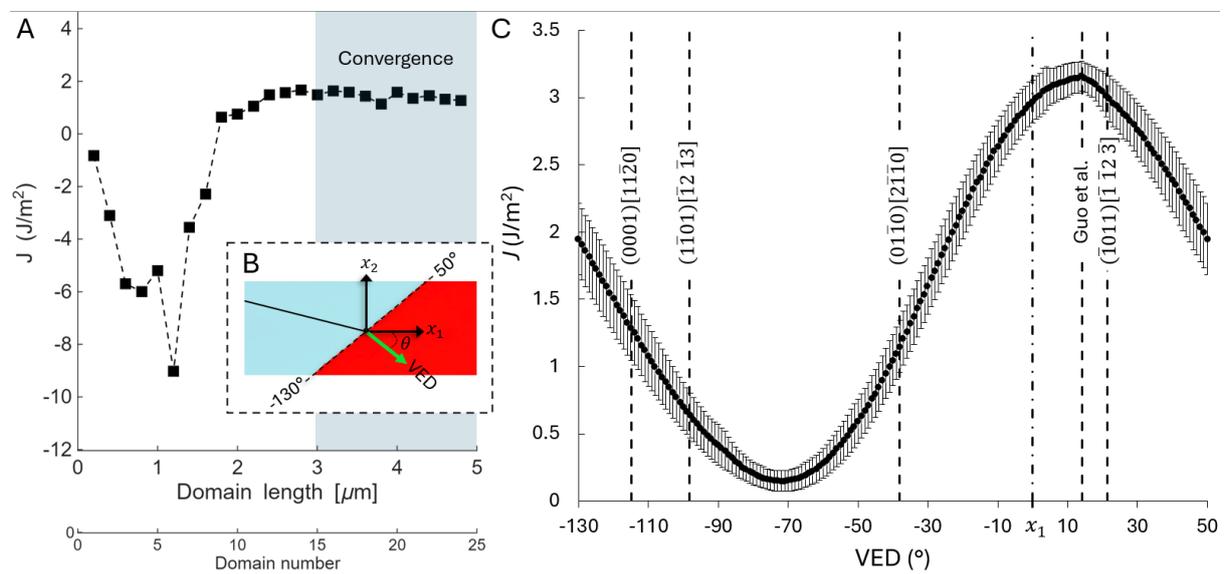

Figure 3: (A) Convergence of configurational force–based *J*-type domain integral on $(01\bar{1}0)[2\bar{1}\bar{1}0]$ prismatic ⟨a⟩ slip with increasing integration domain size. The shaded region highlights the converged regime used for quantitative analysis. (B) The domain integral coordinates superimposed on the rotated EBSD map with the $(01\bar{1}0)[2\bar{1}\bar{1}0]$ slip trace (green arrow) and VED relative to the origin, i.e., stress localisation. (C) Directional dependence of the configurational force evaluated as a function of VED angle, covering the entire probable energy flux direction in Grain B. Vertical dashed lines indicate the projected trace directions of crystallographically admissible slip systems in Grain B, including basal ⟨a⟩, prismatic ⟨a⟩, and pyramidal ⟨c+a⟩. The Guo et al. [19] fitting direction is shown for comparison. Error bars reflect the standard deviation after path-independent convergence.

We assessed the directional nature of the configurational force by rotating displacement-gradient fields into a trace-aligned frame and evaluating the integral versus virtual extension direction along the slip trace (Figure 3B). The resulting response exhibits a clear maximum for a specific projected trace direction corresponding to one of the crystallographically admissible slip systems in Grain B, i.e., $(\bar{1}011)[\bar{1}\,\bar{1}2\bar{3}]$. This demonstrates that the stress localisation induced by slip-band blocking generates a strongly directional driving force that can be quantified without recourse to ad hoc stress fitting or assumed interaction lengths.

A noteworthy result is that the fitting direction selected by Guo et al. [19], when applied to the same dataset, lies close to the direction of maximum configurational force identified in the present analysis. Converting that configurational force to an equivalent stress intensity factor using the Hill-averaged elastic modulus of Grain B (114.62 GPa) yields $K = 0.60 \pm 0.02$ MPa m$^{0.5}$, in close agreement with the value reported by Guo et al. ($0.62 \pm 0.05$ MPa m$^{0.5}$) obtained from $\sigma_{13}$-based fitting after resolving the stress field onto the slip direction. This quantitative agreement is notable given the fundamentally different methodologies. However, it should be emphasised that the configurational force result is not a direct measure of actual slip transfer across the boundary, nor of the intrinsic grain boundary resistance to transfer. Rather, because it is evaluated from the elastic field associated with the measured blocked configuration, it provides an energetic descriptor of the tendency for deformation to extend into Grain B along a given direction. In the present case, that tendency should be interpreted cautiously because Grain B is relatively hard for the incoming deformation mode, so a favourable configurational force direction does not by itself imply that slip transfer is probable.

At the same time, it is essential to recognise that the fitted trace direction reported by Guo et al. does not coincide with any crystallographically admissible slip variant in the neighbouring grain. Additionally, as previously noted by Britton and Wilkinson [20], $\sigma_{13}$-based fitting procedures are highly sensitive to the assumed grain boundary position and to the initial choice of fitting position. The same sensitivity applies to the current analysis, but becomes irrelevant once the domain fully encloses the region of intense stress localisation.

The slip-resolved results demonstrate an apparent decoupling between conventional activation metrics and the energetic driving force generated by the blocked slip band (Table 1). Although the $(01\bar{1}0)[2\bar{1}\bar{1}0]$ prismatic ⟨a⟩ slip variant exhibits the highest Schmid factor (0.49), its associated configurational force is comparatively modest, indicating that crystallographic favourability under the applied load does not directly translate into the strongest local driving force at the grain boundary. Conversely, the $(0001)[11\bar{2}0]$ basal ⟨a⟩ variant exhibits a non-negligible configurational force despite its very low Schmid factor, showing that the configurational force captures aspects of local energy redistribution that are not reflected geometrically. The $(1\bar{1}01)[\bar{1}2\bar{1}3]$ pyramidal ⟨c+a⟩ variant, characterised by both low Schmid factor and low $m'$, is associated with the smallest configurational force, consistent with its unfavourable alignment.

Table 1: Slip variants identified in Grain B, with their type, colour in Figure 2B, Schmid factor, $m'$ and residual Burgers vector ($\vec{b_r}$) calculated relative to $(01\bar{1}0)[2\bar{1}\bar{1}0]$ slip in Grain A, the angle ($\theta$) between the $x_1$-axis and the slip's VED, and the slip-resolved configurational force.

| Slip variant | Type | Colour | SF | $m'$ | $\vec{b_r}$ | $\theta$ (°) | $J$ (J/m²) |
|---|---|---|---|---|---|---|---|
| $(0001)[\bar{1}2\bar{1}0]$ | Basal ⟨a⟩ | Blue | 0.00 | 0.04 | 1.37 | -114.9 | 1.290 ± 0.245 |
| $(0001)[11\bar{2}0]$ | Basal ⟨a⟩ | Blue | 0.01 | 0.43 | 0.84 | -114.9 | 1.290 ± 0.246 |
| $(0001)[2\bar{1}\bar{1}0]$ | Basal ⟨a⟩ | Blue | 0.01 | 0.39 | 0.91 | -114.9 | 1.290 ± 0.247 |
| $(01\bar{1}0)[2\bar{1}\bar{1}0]$ | Prismatic ⟨a⟩ | Yellow | 0.49 | 0.40 | 0.91 | -38.2 | 1.140 ± 0.195 |
| $(1\bar{1}00)[11\bar{2}0]$ | Prismatic ⟨a⟩ | Black | 0.14 | 0.04 | 0.84 | -98.2 | 0.643 ± 0.169 |
| $(\bar{1}010)[\bar{1}2\bar{1}0]$ | Prismatic ⟨a⟩ | Green | 0.35 | 0.04 | 1.37 | 21.8 | 3.000 ± 0.170 |
| $(0\bar{1}1\bar{1})[1\bar{2}13]$ | 1st order pyramidal ⟨c+a⟩ | Yellow | 0.13 | 0.16 | 0.94 | -37.6 | 1.170 ± 0.200 |
| $(0\bar{1}1\bar{1})[11\bar{2}\bar{3}]$ | 1st order pyramidal ⟨c+a⟩ | Yellow | 0.36 | 0.07 | 1.58 | -37.6 | 1.170 ± 0.200 |
| $(01\bar{1}\bar{1})[\bar{1}2\bar{1}3]$ | 1st order pyramidal ⟨c+a⟩ | Yellow | 0.14 | 0.57 | 0.87 | -37.6 | 1.170 ± 0.200 |
| $(01\bar{1}\bar{1})[\bar{1}\bar{1}2\bar{3}]$ | 1st order pyramidal ⟨c+a⟩ | Yellow | 0.37 | 0.85 | 1.97 | -37.6 | 1.170 ± 0.200 |
| $(1\bar{1}01)[\bar{1}2\bar{1}3]$ | 1st order pyramidal ⟨c+a⟩ | Black | 0.03 | 0.23 | 0.87 | -98.3 | 0.646 ± 0.169 |
| $(1\bar{1}01)[\bar{2}113]$ | 1st order pyramidal ⟨c+a⟩ | Black | 0.04 | 0.10 | 1.20 | -98.3 | 0.646 ± 0.169 |
| $(\bar{1}101)[1\bar{2}13]$ | 1st order pyramidal ⟨c+a⟩ | Black | 0.02 | 0.15 | 0.94 | -98.0 | 0.636 ± 0.168 |
| $(\bar{1}101)[2\bar{1}\bar{1}3]$ | 1st order pyramidal ⟨c+a⟩ | Black | 0.04 | 0.24 | 0.43 | -98.0 | 0.636 ± 0.168 |
| $(10\bar{1}1)[11\bar{2}\,\bar{3}]$ | 1st order pyramidal ⟨c+a⟩ | Green | 0.43 | 0.06 | 1.58 | 22.2 | 2.990 ± 0.170 |
| $(10\bar{1}1)[\bar{2}113]$ | 1st order pyramidal ⟨c+a⟩ | Green | 0.26 | 0.06 | 1.20 | 22.2 | 2.990 ± 0.170 |
| $(\bar{1}011)[\bar{1}\,12\bar{3}]$ | 1st order pyramidal ⟨c+a⟩ | Green | 0.43 | 0.80 | 1.97 | 21.4 | 3.010 ± 0.165 |
| $(\bar{1}011)[2\bar{1}\bar{1}3]$ | 1st order pyramidal ⟨c+a⟩ | Green | 0.27 | 0.78 | 0.43 | 21.4 | 3.010 ± 0.165 |
| - | Guo et al. [19] | - | - | - | - | 14.2 | 3.150 ± 0.110 |

While the Schmid factor and geometric slip transferability (m′ and $\vec{b}_r$) remain useful for screening candidate slip systems, they do not uniquely predict the magnitude of the configurational force [44–47]. Here, the residual Burgers vector is included as a measure of the lattice mismatch that would remain at the interface following a putative transfer event, so that smaller values indicate more geometrically compatible transfer pathways. Thus, in this framework, the configurational force identifies the energetic preference for deformation to extend along specific crystallographically admissible directions in Grain B, whereas $m′$ and the residual Burgers vector reflect the geometric compatibility of a possible transfer event. Accordingly, a slip system associated with a large configurational force cannot automatically be regarded as a probable transfer path if it also has an unfavourable residual Burgers vector or belongs to a relatively hard response in Grain B.

In the present case, the results indicate only the $(\bar{1}011)[2\bar{1}\bar{1}3]$ first-order pyramidal ⟨c+a⟩ variant in Grain B is energetically more favourable for slip extension than others; they do not demonstrate actual slip transmission across the boundary especially due to the large residual Burgers vector. Hence, since Grain B is a relatively hard grain for the incoming slip, the observed localisation is more consistent with continued blockage and stress accumulation than with straightforward slip transmission, even where the configurational force analysis indicates an energetically favourable extension direction. Therefore, the present results should be interpreted as identifying the energetic tendency for deformation to extend into Grain B, conditioned by $m′$ and the residual Burgers vector, rather than as evidence that slip transfer has actually occurred or is necessarily probable in this hard-grain configuration.

Treating the blocked slip-band tip within a configurational force framework provides a scalar, coordinate-invariant measure of the local energetic driving force available to promote further deformation in the vicinity of the grain boundary, including possible extension into the neighbouring grain. This is particularly as a slip-induced localisation does not necessarily relax by transmitting slip into the neighbouring grain, but can lead to cavity or crack initiations [15,48–50], as from a continuum perspective, and even in the absence of a physical crack or cavity, this region may be interpreted as a confined micro-damage precursor, commonly described in the literature as a "microvolume" or "blooming zone" [2,51,52]. Such zones are characterised by near-singular stress and strain fields and high geometrically necessary dislocation density with reduced capacity for further dislocation motion that intensifies with load, and can therefore act as an efficient sink for the elastic energy accumulated ahead of the blocked slip-band [53]. Thus, quantifying these localised energy concentrations via configurational force integrals provides a physically grounded route to assessing damage initiation prior to the formation of discrete cracks.

The present method ranks energetically favourable extension directions for the measured blocked state, but it does not by itself provide a criterion for whether transfer will occur. More fundamentally, the present analysis does not establish the critical energetic condition required for actual slip transfer. A more suitable objective would be to determine the energy required for transfer, for example, through a critical configurational force or critical $J_c$-type threshold, but this cannot be inferred from a single postmortem blocked-slip configuration. Establishing such a threshold would require in situ experiments capable of resolving the onset of transfer directly, so that the critical state at which a blocked slip band either transmits, remains arrested, or nucleates damage can be identified quantitatively. Therefore, to capture the full deformation state, future work should use the current analysis with full-field displacement-based measurements, such as scanning electron microscopy digital image correlation, coupled with crystal plasticity to calculate plastic stresses [54,55].

## 4. Conclusion

In summary, this work demonstrates that configurational force analysis provides a robust, energy-based descriptor of the severity and directionality of slip-band blocking at grain boundaries. Unlike conventional geometric metrics, the approach parameterises the measured elastic field and identifies the directions in which deformation extension into the neighbouring grain is energetically more favourable. However, the configurational force result should not be interpreted as a direct measure of actual slip transfer or of the intrinsic resistance of the boundary to transfer. In the present case, where Grain B behaves as a relatively hard grain for the incoming slip, the analysis supports an energetic tendency for extension into Grain B but does not establish that slip transfer is probable. A more complete treatment of transfer would require the determination of a critical transfer threshold, likely through in situ experiments capable of defining a critical $J_c$.

# Acknowledgement

We thank Professor Angus J. Wilkinson and Dr Phani Karamched (University of Oxford) for sharing the Ti HR-EBSD data, and Dr Zachary Kloenne (Imperial College London) for discussion on titanium. Financial support from the Royal Society via the Short Industry Fellowships (SIF\R2\242005) are gratefully acknowledged.

# References

[1]    Z. Zheng, D.S. Balint, F.P.E. Dunne, Investigation of slip transfer across HCP grain boundaries with application to cold dwell facet fatigue, Acta Mater. 127 (2017) 43–53. https://doi.org/10.1016/j.actamat.2017.01.021.

[2]    J.C. Stinville, M.A. Charpagne, R. Maaß, H. Proudhon, W. Ludwig, P.G. Callahan, F. Wang, I.J. Beyerlein, M.P. Echlin, T.M. Pollock, Insights into Plastic Localization by Crystallographic Slip from Emerging Experimental and Numerical Approaches, Annu. Rev. Mater. Res. 53 (2023) 275–317. https://doi.org/10.1146/annurev-matsci-080921-102621.

[3]    G.A. Malygin, Plasticity and strength of micro- and nanocrystalline materials, Physics of the Solid State 49 (2007) 1013–1033. https://doi.org/10.1134/S1063783407060017.

[4]    J.P. Hirth, The influence of grain boundaries on mechanical properties, Metallurgical Transactions 3 (1972) 3047–3067. https://doi.org/10.1007/BF02661312.

[5]    P.M. Anderson, J.P. Hirth, J. Lothe, Theory of Straight Dislocations, in: Theory of Dislocations, Third, Cambridge University Press, New York, 2017: pp. 85–89.

[6]    Y. Su, T. Phan, L. Xiong, J. Kacher, Multiscale computational and experimental analysis of slip-GB reactions: In situ high-resolution electron backscattered diffraction and concurrent atomistic-continuum simulations, Scr. Mater. 232 (2023) 115500. https://doi.org/10.1016/j.scriptamat.2023.115500.

[7]    J.C. Stinville, J.M. Hestroffer, M.A. Charpagne, A.T. Polonsky, M.P. Echlin, C.J. Torbet, V. Valle, K.E. Nygren, M.P. Miller, O. Klaas, A. Loghin, I.J. Beyerlein, T.M. Pollock, Multi-

modal Dataset of a Polycrystalline Metallic Material: 3D Microstructure and Deformation Fields, Sci. Data 9 (2022) 460. https://doi.org/10.1038/s41597-022-01525-w.

[8]     B. Larrouy, P. Villechaise, J. Cormier, O. Berteaux, Grain boundary–slip bands interactions: Impact on the fatigue crack initiation in a polycrystalline forged Ni-based superalloy, Acta Mater. 99 (2015) 325–336. https://doi.org/10.1016/j.actamat.2015.08.009.

[9]     M.I. Latypov, J.M. Hestroffer, J.-C. Stinville, J.R. Mayeur, T.M. Pollock, I.J. Beyerlein, Modeling lattice rotation fields from discrete crystallographic slip bands in superalloys, Extreme Mech. Lett. 49 (2021) 101468. https://doi.org/10.1016/j.eml.2021.101468.

[10]    Z. Zhang, D. Lunt, H. Abdolvand, A.J. Wilkinson, M. Preuss, F.P.E. Dunne, Quantitative investigation of micro slip and localization in polycrystalline materials under uniaxial tension, Int. J. Plast. 108 (2018) 88–106. https://doi.org/10.1016/J.IJPLAS.2018.04.014.

[11]    E. Bayerschen, A.T. McBride, B.D. Reddy, T. Böhlke, Review on slip transmission criteria in experiments and crystal plasticity models, J. Mater. Sci. 51 (2016) 2243–2258. https://doi.org/10.1007/s10853-015-9553-4.

[12]    Y. Su, S. Han, P. Eisenlohr, M.A. Crimp, Predicting shear transmission across grain boundaries with an iterative stress relief model, Acta Mater. 215 (2021) 116992. https://doi.org/10.1016/j.actamat.2021.116992.

[13]    T.R. Bieler, P. Eisenlohr, F. Roters, D. Kumar, D.E. Mason, M.A. Crimp, D. Raabe, The role of heterogeneous deformation on damage nucleation at grain boundaries in single phase metals, Int. J. Plast. 25 (2009) 1655–1683. https://doi.org/10.1016/j.ijplas.2008.09.002.

[14]    L. Wang, Y. Yang, P. Eisenlohr, T.R. Bieler, M.A. Crimp, D.E. Mason, Twin Nucleation by Slip Transfer across Grain Boundaries in Commercial Purity Titanium, Metallurgical and Materials Transactions A 41 (2009) 421. https://doi.org/10.1007/s11661-009-0097-6.

[15]  M.H. Yoo, H. Trinkaus, Interaction of slip with grain boundary and its role in cavity nucleation, Acta Metallurgica 34 (1986) 2381–2390. https://doi.org/10.1016/0001-6160(86)90141-0.

[16]  R.L. Black, D. Anjaria, J. Genée, V. Valle, J.C. Stinville, Micro-strain and cyclic slip accumulation in a polycrystalline nickel-based superalloy, Acta Mater. 266 (2024) 119657. https://doi.org/10.1016/j.actamat.2024.119657.

[17]  A.J. Wilkinson, G. Meaden, D.J. Dingley, High resolution mapping of strains and rotations using electron backscatter diffraction, Materials Science and Technology 22 (2006) 1271–1278. https://doi.org/10.1179/174328406X130966.

[18]  T.J. Ruggles, W.G. Gilliland, D.T. Fullwood, J. Kacher, An overview of HR-EBSD techniques for mapping local stress and dislocations in crystalline materials at sub-micron resolution, Prog. Mater. Sci. 157 (2026) 101585. https://doi.org/10.1016/j.pmatsci.2025.101585.

[19]  Y. Guo, T.B. Britton, A.J. Wilkinson, Slip band–grain boundary interactions in commercial-purity titanium, Acta Mater. 76 (2014) 1–12. https://doi.org/10.1016/J.ACTAMAT.2014.05.015.

[20]  T.B. Britton, A.J. Wilkinson, Stress fields and geometrically necessary dislocation density distributions near the head of a blocked slip band, Acta Mater. 60 (2012) 5773–5782. https://doi.org/10.1016/j.actamat.2012.07.004.

[21]  J.D. Eshelby, F.C. Frank, F.R.N. Nabarro, XLI. The equilibrium of linear arrays of dislocations., The London, Edinburgh, and Dublin Philosophical Magazine and Journal of Science 42 (1951) 351–364. https://doi.org/10.1080/14786445108561060.

[22]  B. Guan, Y. Xin, X. Huang, C. Liu, P. Wu, Q. Liu, The mechanism for an orientation dependence of grain boundary strengthening in pure titanium, Int. J. Plast. 153 (2022) 103276. https://doi.org/10.1016/j.ijplas.2022.103276.

[23]  M.T. Andani, A. Lakshmanan, V. Sundararaghavan, J. Allison, A. Misra, Quantitative study of the effect of grain boundary parameters on the slip system level Hall-Petch

slope for basal slip system in Mg-4Al, Acta Mater. 200 (2020) 148–161. https://doi.org/10.1016/j.actamat.2020.08.079.

[24] M.T. Andani, A. Lakshmanan, M. Karamooz-Ravari, V. Sundararaghavan, J. Allison, A. Misra, A quantitative study of stress fields ahead of a slip band blocked by a grain boundary in unalloyed magnesium, Sci. Rep. 10 (2020) 3084. https://doi.org/10.1038/s41598-020-59684-y.

[25] C. Fressengeas, M. V. Upadhyay, A continuum model for slip transfer at grain boundaries, Adv. Model. Simul. Eng. Sci. 7 (2020) 12. https://doi.org/10.1186/s40323-020-00145-6.

[26] J.D. Eshelby, The elastic energy-momentum tensor, J. Elast. 5 (1975) 321–335. https://doi.org/10.1007/BF00126994.

[27] J.D. Eshelby, The Continuum Theory of Lattice Defects, in: F. Seitz, D.B.T.-S.S.P. Turnbull (Eds.), Academic Press, 1956: pp. 79–144. https://doi.org/10.1016/S0081-1947(08)60132-0.

[28] T.B. Britton, A.J. Wilkinson, High resolution electron backscatter diffraction measurements of elastic strain variations in the presence of larger lattice rotations, Ultramicroscopy 114 (2012) 82–95. https://doi.org/10.1016/j.ultramic.2012.01.004.

[29] C. Maurice, J.H. Driver, R. Fortunier, On solving the orientation gradient dependency of high angular resolution EBSD, Ultramicroscopy 113 (2012) 171–181. https://doi.org/10.1016/j.ultramic.2011.10.013.

[30] F.P.E. Dunne, D. Rugg, A. Walker, Lengthscale-dependent, elastically anisotropic, physically-based hcp crystal plasticity: Application to cold-dwell fatigue in Ti alloys, Int. J. Plast. 23 (2007) 1061–1083. https://doi.org/10.1016/j.ijplas.2006.10.013.

[31] T.J. HARDIN, T.J. RUGGLES, D.P. KOCH, S.R. NIEZGODA, D.T. FULLWOOD, E.R. HOMER, Analysis of traction-free assumption in high-resolution EBSD measurements, J. Microsc. 260 (2015) 73–85. https://doi.org/10.1111/jmi.12268.

[32]    V.A. Lubarda, The energy momentum tensor in the presence of body forces and the Peach–Koehler force on a dislocation, Int. J. Solids Struct. 45 (2008) 1536–1545. https://doi.org/10.1016/j.ijsolstr.2007.10.004.

[33]    J.R. Rice, A Path Independent Integral and the Approximate Analysis of Strain Concentration by Notches and Cracks, J. Appl. Mech. 35 (1968) 379–386. https://doi.org/10.1115/1.3601206.

[34]    F.Z. Li, C.F. Shih, A. Needleman, A comparison of methods for calculating energy release rates, Eng. Fract. Mech. 21 (1985) 405–421. https://doi.org/10.1016/0013-7944(85)90029-3.

[35]    A. Koko, A. Abdelnour, T.H. Becker, T.J. Marrow, Bridging experiments and defects' mechanics: a data-driven toolbox for configurational force analysis, Eng. Comput. 42 (2026) 21. https://doi.org/10.1007/s00366-025-02262-5.

[36]    G.P. Nikishkov, A.V. Vershinin, Y.G. Nikishkov, Mesh-independent equivalent domain integral method for J-integral evaluation, Advances in Engineering Software 100 (2016) 308–318. https://doi.org/10.1016/j.advengsoft.2016.08.006.

[37]    A. Koko, T.H. Becker, E. Elmukashfi, N.M. Pugno, A.J. Wilkinson, T.J. Marrow, HR-EBSD analysis of in situ stable crack growth at the micron scale, J. Mech. Phys. Solids 172 (2023) 105173. https://doi.org/10.1016/j.jmps.2022.105173.

[38]    D.M. Parks, The virtual crack extension method for nonlinear material behavior, Comput. Methods Appl. Mech. Eng. 12 (1977) 353–364. https://doi.org/10.1016/0045-7825(77)90023-8.

[39]    S.G. Lekhnitskii, General Equations of the Theory of Elasticity of an Anisotropic Body, in: Theory of Elasticity of an Anisotropic Elastic Body, 1st ed., MIR Publishers, Moscow, 1981: pp. 15–73. https://archive.org/details/lekhnitskii-theory-of-elasticity-of-an-anisotropic-body-mir-1981/page/6/mode/2up (accessed October 25, 2021).

[40]    D. Mainprice, F. Bachmann, R. Hielscher, H. Schaeben, Descriptive tools for the analysis of texture projects with large datasets using MTEX : strength, symmetry and

components, Geological Society, London, Special Publications 409 (2015) 251–271. https://doi.org/10.1144/SP409.8.

[41] R. Zhang, Q. Zhao, Y. Zhao, D. Guo, Y. Du, Research Progress on Slip Behavior of α-Ti under Quasi-Static Loading: A Review, Metals (Basel). 12 (2022) 1571. https://doi.org/10.3390/met12101571.

[42] M. Baral, T. Hama, E. Knudsen, Y.P. Korkolis, Plastic deformation of commercially-pure titanium: experiments and modeling, Int. J. Plast. 105 (2018) 164–194. https://doi.org/10.1016/j.ijplas.2018.02.009.

[43] K. Hamulka, T. Vermeij, A. Sharma, R. Pero, J. Michler, X. Maeder, Effects of strain rate and c-axis orientation on microscale α-Ti compression: From kink bands to twinning, Acta Mater. (2026) 121923. https://doi.org/10.1016/j.actamat.2026.121923.

[44] T.R. Bieler, P. Eisenlohr, C. Zhang, H.J. Phukan, M.A. Crimp, Grain boundaries and interfaces in slip transfer, Curr. Opin. Solid State Mater. Sci. 18 (2014) 212–226. https://doi.org/10.1016/J.COSSMS.2014.05.003.

[45] D. Mercier, C. Zambaldi, T.R. Bieler, A Matlab toolbox to analyze slip transfer through grain boundaries, IOP Conf. Ser. Mater. Sci. Eng. 82 (2015) 12090. https://doi.org/10.1088/1757-899x/82/1/012090.

[46] T.R. Bieler, R. Alizadeh, M. Peña-Ortega, J. Llorca, An analysis of (the lack of) slip transfer between near-cube oriented grains in pure Al, Int. J. Plast. 118 (2019) 269–290. https://doi.org/10.1016/j.ijplas.2019.02.014.

[47] E. Nieto-Valeiras, A. Orozco-Caballero, M. Sarebanzadeh, J. Sun, J. LLorca, Analysis of slip transfer across grain boundaries in Ti via diffraction contrast tomography and high-resolution digital image correlation: When the geometrical criteria are not sufficient, Int. J. Plast. 175 (2024) 103941. https://doi.org/10.1016/j.ijplas.2024.103941.

[48] K.S. Stopka, M. Yaghoobi, J.E. Allison, D.L. McDowell, Effects of Boundary Conditions on Microstructure-Sensitive Fatigue Crystal Plasticity Analysis, Integr. Mater. Manuf. Innov. 10 (2021) 393–412. https://doi.org/10.1007/s40192-021-00219-2.

[49]   X. Cao, L. Zhang, H. Duan, Z. Zhou, J. Qiu, C. Fang, B. Jiang, J. Lei, R. Yang, K. Du, The slip transfer effect on crack nucleation under dwell fatigue loading in titanium alloys, J. Mater. Sci. 60 (2025) 20675–20688. https://doi.org/10.1007/s10853-025-10963-x.

[50]   M.D. Sangid, H.J. Maier, H. Sehitoglu, The role of grain boundaries on fatigue crack initiation – An energy approach, Int. J. Plast. 27 (2011) 801–821. https://doi.org/10.1016/j.ijplas.2010.09.009.

[51]   D.J. Shadle, K.E. Nygren, J.C. Stinville, M.A. Charpagne, T.J.H. Long, M.P. Echlin, C.J. Budrow, A.T. Polonsky, T.M. Pollock, I.J. Beyerlein, M.P. Miller, Integrating in-situ multi-modal characterizations with signatures to investigate localized deformation, Mater. Charact. 205 (2023) 113332. https://doi.org/10.1016/j.matchar.2023.113332.

[52]   S. Sun, B.L. Adams, W.E. King, Observations of lattice curvature near the interface of a deformed aluminium bicrystal, Philosophical Magazine A 80 (2000) 9–25. https://doi.org/10.1080/01418610008212038.

[53]   A. Koko, In situ characterisation of slip-band behaviour in ferrite under mechanical loading, Proceedings of the Royal Society A: Mathematical, Physical and Engineering Sciences 481 (2025). https://doi.org/10.1098/rspa.2023.0804.

[54]   A. Koko, B. Sheen, C. Green, F. Dunne, In situ elucidation of fracture mechanisms governing crack transition to plasticity arrest, International Journal of Mechanical Sciences (under Review) (2026). http://arxiv.org/abs/2510.03713.

[55]   D. Depriester, J.P. Goulmy, L. Barrallier, Crystal Plasticity simulations of in situ tensile tests: A two-step inverse method for identification of CP parameters, and assessment of CPFEM capabilities, Int. J. Plast. 168 (2023) 103695. https://doi.org/10.1016/j.ijplas.2023.103695.